\documentclass{article}
\usepackage{PRIMEarxiv}
\usepackage[utf8]{inputenc} 
\usepackage[T1]{fontenc}    
\usepackage[hidelinks]{hyperref}       
\usepackage{url}            
\usepackage{booktabs}       
\usepackage{amssymb,amsfonts,amsmath}       
\usepackage{nicefrac}       
\usepackage{microtype}      
\usepackage{tabularx}
\usepackage{longtable}
\usepackage{graphicx}
\graphicspath{{media/}}

\title{On the Time-Domain Waveforms Leading to Frequency Comb Spectra}

\author{
  Madhurendra Mishra$^{1}$ \thanks{ORCID: 0009-0003-0080-6186} \\
  Department of Physics \\
  Sri Guru Tegh Bahadur Khalsa College, University of Delhi \\
  New Delhi -- 110007, India \\
  \texttt{madhurendramishra24@gmail.com} \\
  \And
  Adarsh Ganesan$^{2}$ \thanks{Corresponding author. ORCID: 0000-0002-5107-8452} \\
  Department of Electrical and Electronics Engineering \\
  Birla Institute of Technology and Science Pilani, Dubai Campus \\
  Dubai International Academic City, Dubai -- 345055, UAE \\
  \texttt{adarsh@dubai.bits-pilani.ac.in} \\
}

\begin{document}
\maketitle

\begin{abstract}
Frequency-comb spectra can arise from distinct time-domain waveforms. We develop a Fourier-based framework that relates waveform symmetry, regularity, and modulation to the structure of overtone- and sideband frequency combs. Analytical derivations show that symmetry determines the allowed harmonic orders, while waveform regularity governs spectral decay: discontinuities produce $n^{-1}$ scaling, derivative discontinuities yield $n^{-2}$ scaling, and smooth nonlinear mappings can produce rapidly decaying harmonics. For modulated carriers, we distinguish consecutive sidebands, symmetry-induced missing orders, and nonuniformly spaced frequency components. In particular, spectral sparsity does not necessarily imply a nonuniform frequency lattice. Comparisons of analytical coefficients with calculated time-domain signals and Fourier spectra establish the correspondence between waveform properties and spectral signatures. The results provide a mathematical basis for interpreting frequency-comb structures while identifying the limitations of magnitude-only spectra in determining their underlying physical mechanisms.
\end{abstract}


\section{Introduction}

Optical frequency combs provide a set of equally spaced spectral lines with stable frequency and phase relationships, allowing optical frequencies to be measured and compared with exceptional precision. They have become established tools in frequency metrology, optical clocks, and spectroscopy \cite{udem2002optical,cundiff2003colloquium,fortier2019twenty}. Similar spectral structures can arise in mechanical resonators, although the underlying processes need not be the same. In a nonlinear mechanical system, a comb may reflect modulation of a driven response, coupling between vibrational modes, or the harmonic content of the motion itself.

The early development of phononic frequency combs was closely tied to nonlinear resonances and frequency mixing \cite{mahboob2012tuneable,cao2014phononic}. Ganesan et al. demonstrated comb formation through intrinsic three-wave mixing between a driven mechanical mode and an autoparametrically excited mode \cite{ganesan2017phononic}. Subsequent work explored four-wave mixing, three-mode parametric resonance, and comb formation near bifurcations associated with internal resonance \cite{ganesan2017fourwave,ganesan2018threemode,czaplewski2018bifurcation,seitner2017parametric}. Related responses have been reported in bulk acoustic resonators, single-mode nanomechanical systems, symmetry-broken graphene resonators, and parametrically driven Duffing oscillators \cite{goryachev2020generation,ochs2022single,keskekler2022symmetry,batista2020frequency}. These studies established several routes to the closely spaced sidebands commonly associated with phononic combs, while making clear that mode coupling is not a universal requirement.

Mechanical overtone frequency combs provide a different route. De Jong et al. generated up to 35 harmonics of a single mechanical oscillation by coupling a suspended membrane to a spatially periodic optical field \cite{dejong2023mechanical}. The position-dependent optical force produced spectral components at integer multiples of the fundamental frequency, with fixed frequency and phase relationships. Rather than requiring a sequence of equally spaced mechanical eigenmodes, the comb arose from the nonlinear response of one oscillating structure. This distinction is important: overtones follow the fundamental frequency, whereas mixing-generated sidebands are spaced around driven or internally excited frequencies. The two mechanisms can also coexist. Multi-harmonic mechanical combs and hybrid optomechanical and electromechanical combs have been investigated in several platforms \cite{bhosale2024multiharmonic,miri2018optomechanical,wu2022hybridized}. More recently, Xiao et al. reported an ultrabroadband phonon-laser comb with more than 6,000 spectral lines, where phonon-lasing harmonics coexist with sidebands produced by cascaded four-wave mixing \cite{xiao2026ultrabroadband}. Such spectra need not be attributed to merely one nonlinear process; their breadth can reflect the combined action of harmonic generation and frequency mixing.

While different periodic waveforms admit a harmonic expansion, the distribution of spectral weight depends on its shape, and the power spectrum alone does not retain the relative phases needed to reconstruct that waveform. Here, we establish a Fourier-based framework that connects time-domain waveforms to the spectral structure of frequency combs. We examine mathematical constructions encompassing overtone harmonics, sidebands, and non-equidistant frequency spectra. By deriving their Fourier coefficients and finite-window transforms, we determine how waveform symmetry, regularity, and modulation govern harmonic selection, spectral decay, and frequency spacing.

\section{Mathematical Structures}\label{sec:math}

\subsection{Fourier representation and spectral regularity}\label{subsec:fourier-framework}

For a real, $T$-periodic signal $x(t)$, let $\omega_0=2\pi/T$ and
\begin{equation}\label{eq:fourier}
 x(t)=\sum_{n\in\mathbb Z}c_n e^{in\omega_0t},\qquad
 c_n=\frac{1}{T}\int_0^T x(t)e^{-in\omega_0t}\,dt.
\end{equation}
Since $c_{-n}=c_n^*$, the positive-frequency representation is
\begin{equation}\label{eq:amplitude}
 x(t)=c_0+\sum_{n=1}^{\infty}
 A_n\cos(n\omega_0t+\phi_n),\qquad
 A_n=2|c_n|,\quad \phi_n=\arg c_n.
\end{equation}
Here $A_n$ is a peak amplitude, not a power; Parseval's identity assigns $A_n^2/2$ to its contribution to the mean square. The translation of time modifies $\phi_n$ but leaves $A_n$ invariant. Consequently, a magnitude spectrum does not uniquely determine the waveform.

For a duration centered observation interval $T_{\rm obs}$, the frequency-domain transform is
\begin{align}\label{eq:finite-window}
 X_{\rm obs}(\omega)&=\int_{-T_{\rm obs}/2}^{T_{\rm obs}/2}
 x(t)e^{-i\omega t}\,dt
 =\sum_{n\in\mathbb Z}c_n W(\omega-n\omega_0),\\
 W(\nu)&=T_{\rm obs}\,\operatorname{sinc}\!\left(\frac{\nu T_{\rm obs}}{2}\right),
 \quad\operatorname{sinc}(z)=\frac{\sin z}{z},\notag
\end{align}
with $\operatorname{sinc}(0)=1$. The ideal infinite-duration transform is $X(\omega)=2\pi\sum_n c_n\delta(\omega-n\omega_0)$. The formulas below give $X_{\rm obs}(\omega)$ for the signals already defined in the section; the ideal line spectrum follows by replacing $W(\nu)$ with $2\pi\delta(\nu)$.

Spectral selection and decay are governed by different properties. The relation $x(t+T/2)=-x(t)$ implies
\begin{equation}\label{eq:half-wave}
 c_n=(-1)^{n+1}c_n,\qquad c_{2m}=0,
\end{equation}
whereas a function with period $T/2$ has only even Fourier indices when expanded relative to $T$. For a piecewise smooth signal with jumps $[x]_j=x(t_j^+)-x(t_j^-)$, integration by parts gives, under piecewise $C^2$ regularity,
\begin{equation}\label{eq:jump-asymptotic}
 c_n=\frac{1}{in\omega_0T}
 \sum_j[x]_j e^{-in\omega_0t_j}+O(n^{-2}).
\end{equation}
Thus jumps generally produce an $n^{-1}$ amplitude envelope; if $x$ is continuous and its first derivative has jumps, the leading decay is instead $O(n^{-2})$. Cancellations may eliminate selected orders. The waveforms below are prescribed mathematical signals, not assumed solutions of a mechanical equation of motion.

\subsection{Overtone Frequency Combs}\label{subsec:overtones}

\subsubsection{Square wave}\label{subsec:square}

For $\theta=\omega_0t$, take $x_{\rm sq}(t)=A\,\operatorname{sgn}(\sin\theta)$, where $A>0$. Integrating over the positive and negative half-cycles yields
\begin{align}
 c_n^{\rm sq}
 &=\frac{A}{2\pi}\left(\int_0^\pi-\int_{-\pi}^0\right)
 e^{-in\theta}\,d\theta\notag\\
 &=\begin{cases}
 \dfrac{2A}{i\pi n},&n\ \text{odd},\\[3pt]
 0,&n\ \text{even},
 \end{cases}\label{eq:square-coeff}\\
 x_{\rm sq}(t)
 &=\frac{4A}{\pi}\sum_{m=0}^{\infty}
 \frac{\sin[(2m+1)\theta]}{2m+1}.
 \label{eq:square-series}
\end{align}
\begin{equation}\label{eq:square-frequency}
 X_{\rm obs}^{\rm sq}(\omega)=\frac{2A}{i\pi}
 \sum_{m=0}^{\infty}\frac{1}{2m+1}
 \bigl[W(\omega-(2m+1)\omega_0)
       -W(\omega+(2m+1)\omega_0)\bigr].
\end{equation}
The mean is zero and the nonzero peak amplitudes are $A_n=4A/(\pi n)$, $n$ odd. Half-wave antisymmetry suppresses even orders, while the jumps determine the $n^{-1}$ tail. A DC offset affects only $c_0$; unequal pulse widths remove the odd-harmonic restriction. At each discontinuity the Fourier series converges to the midpoint of the one-sided limits.

\subsubsection{Sawtooth wave}\label{subsec:sawtooth}

Define the periodically continued ascending ramp $x_{\rm saw}(t)=A\theta/\pi$ for $-\pi<\theta<\pi$. Its sine coefficients are
\begin{equation}\label{eq:saw}
 b_n=\frac{2A}{\pi^2}\int_0^\pi\theta\sin(n\theta)\,d\theta
 =\frac{2A(-1)^{n+1}}{\pi n},\qquad
 x_{\rm saw}(t)=\sum_{n=1}^{\infty}b_n\sin(n\theta).
\end{equation}
\begin{equation}\label{eq:saw-frequency}
 X_{\rm obs}^{\rm saw}(\omega)=\frac{A}{i\pi}
 \sum_{n=1}^{\infty}\frac{(-1)^{n+1}}{n}
 \bigl[W(\omega-n\omega_0)-W(\omega+n\omega_0)\bigr].
\end{equation}
All positive orders occur, with $A_n=2A/(\pi n)$. The alternating signs fix their relative phases. Unlike the square wave, the sawtooth has no half-wave antisymmetry; its periodic reset nevertheless produces the same $n^{-1}$ decay.

\subsubsection{Periodic pulse train}\label{subsec:pulse-train}

Consider a pulse of height $A$ and width $\tau$ centred in each period $T$, with duty cycle $d=\tau/T\in(0,1)$. For $x_{\rm p}(t)=A$ when $|t|<\tau/2$ modulo $T$, and zero otherwise,
\begin{equation}\label{eq:pulse-coeff}
 c_n^{\rm p}=\frac{A}{T}\int_{-\tau/2}^{\tau/2}
 e^{-in\omega_0t}\,dt
 =Ad\,\operatorname{sinc}(\pi nd),
\end{equation}
\begin{equation}\label{eq:pulse-frequency}
 X_{\rm obs}^{\rm p}(\omega)=Ad\sum_{n\in\mathbb Z}
 \operatorname{sinc}(\pi nd)W(\omega-n\omega_0).
\end{equation}
Here $\operatorname{sinc}z=\sin z/z$ and $\operatorname{sinc}0=1$. Hence $c_0=Ad$ and $A_n=2Ad|\operatorname{sinc}(\pi nd)|$. The continuous envelope has its first zero at $n=1/d$; exact Fourier zeros satisfy $nd\in\mathbb Z\setminus\{0\}$. Decreasing $d$ at fixed pulse height broadens the envelope but reduces $c_0$ and the low-order amplitudes. In particular, $d=1/2$ suppresses nonzero even harmonics. Pulse translation changes coefficient phases but not magnitudes.

\subsubsection{Nonlinear functions of a sinusoid}\label{subsec:nonlinear}

Let $q(t)=B\cos\theta$ and $x(t)=F(q(t))$, where $F$ is a real, instantaneous transfer function. Because $F(B\cos\theta)$ is even in $\theta$,
\begin{equation}\label{eq:general-nonlinear-map}
 c_n=\frac{1}{\pi}\int_0^\pi
 F(B\cos\theta)\cos(n\theta)\,d\theta.
\end{equation}
\begin{equation}\label{eq:nonlinear-frequency}
 X_{\rm obs}^{F}(\omega)=\frac{1}{\pi}\sum_{n\in\mathbb Z}
 W(\omega-n\omega_0)\!
 \int_0^\pi F(B\cos\theta)\cos(n\theta)\,d\theta.
\end{equation}
The substitution $\theta\mapsto\theta+\pi$ shows that an even $F$ produces only even indices, whereas an odd $F$ produces only odd indices. These restrictions do not require a nonsinusoidal input. Explicit polynomial, exponential, and rectification maps are evaluated in Secs.~\ref{subsubsec:polynomial}, \ref{subsubsec:exponential}, and \ref{subsubsec:rectified}. They may describe a nonlinear readout rather than the displacement itself.

\subsubsection{Periodic parabolic waveform and Bernoulli harmonics}\label{subsec:periodic-parabolic}

A different mathematical overtone family is obtained from a periodic
parabolic segment, rather than from a prescribed nonlinear oscillator.
With $\theta=\omega_0t$, define the wrapped phase
$u(\theta)=((\theta+\pi)\bmod 2\pi)-\pi$ and let
\begin{equation}\label{eq:parabolic-definition}
 x_{\rm B}(t)=A\left[\frac{u(\theta)^2}{\pi^2}-\frac13\right].
\end{equation}
On $-\pi\leq\theta<\pi$, this is a centred parabola; periodic
continuation makes it a continuous function with slope discontinuities
at odd multiples of $\pi$. It is four times the periodically extended
second Bernoulli polynomial, evaluated at $(\theta+\pi)/(2\pi)$.
Its Fourier coefficients follow by integrating the quadratic segment
\cite{NISTDLMF}:
\begin{equation}\label{eq:parabolic-coeff}
 c_0^{\rm B}=0,\qquad
 c_{\pm n}^{\rm B}=\frac{2A(-1)^n}{\pi^2n^2},\quad n\geq1.
\end{equation}
Equivalently, the real Fourier series is
\begin{equation}\label{eq:parabolic-series}
 x_{\rm B}(t)=\frac{4A}{\pi^2}
 \sum_{n=1}^{\infty}\frac{(-1)^n}{n^2}
 \cos(n\omega_0t).
\end{equation}
Under the finite observation window of Eq.~\eqref{eq:finite-window},
\begin{align}\label{eq:parabolic-frequency}
 X_{\rm obs}^{\rm B}(\omega)=\frac{2A}{\pi^2}
 \sum_{n=1}^{\infty}\frac{(-1)^n}{n^2}
 \bigl[&W(\omega-n\omega_0)\notag\\
       +{}&W(\omega+n\omega_0)\bigr].
\end{align}
All positive integer harmonic orders occur with peak amplitudes
$A_n=4A/(\pi^2n^2)$. The $n^{-2}$ envelope reflects the
first-derivative discontinuity at the periodic join. Unlike the
triangular wave below, this construction has no odd-only selection
rule. It is a mathematical periodic signal, not a claimed trajectory
of any particular mechanical model.

\begin{figure}
    \centering
    \includegraphics[width=\linewidth]{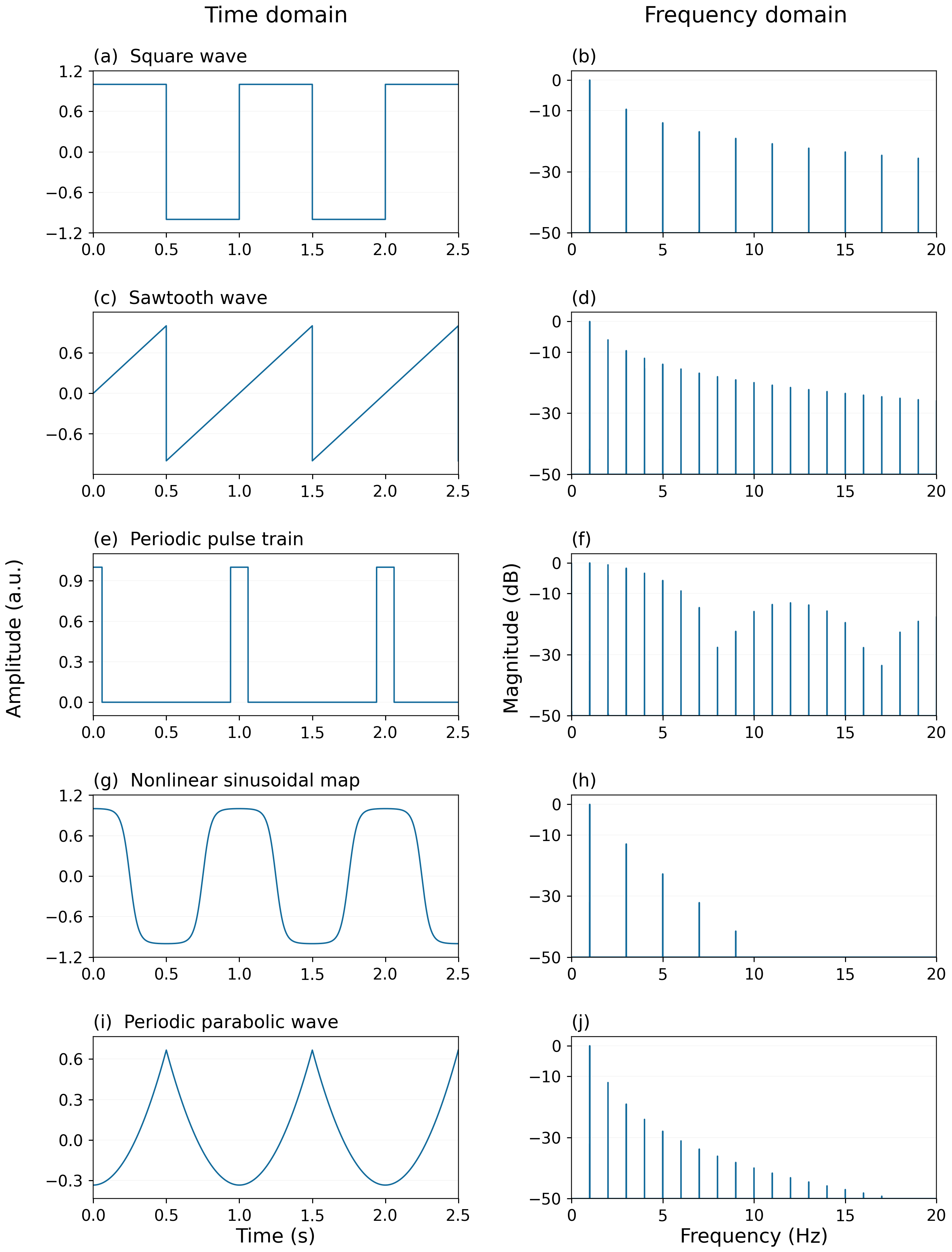}
    \caption{Time-domain waveforms (left column) and their corresponding Fourier magnitude spectra (right column) for five mathematical constructions of overtone frequency combs. Panels~(a,b): square wave; panels~(c,d): sawtooth wave; panels~(e,f): periodic rectangular pulse train; panels~(g,h): nonlinear sinusoidal map; and panels~(i,j): periodic parabolic wave.}
    \label{fig1}
\end{figure}

\subsubsection{Triangular wave}\label{subsubsec:triangle}

The zero-mean triangular wave with extrema $\pm A$ is
\begin{equation}\label{eq:triangle-definition}
 x_{\rm tri}(t)=\frac{2A}{\pi}\arcsin(\sin\theta).
\end{equation}
Its derivative away from the corners is $(2A\omega_0/\pi)\operatorname{sgn}(\cos\theta)$. Integrating the corresponding square-wave expansion gives
\begin{equation}\label{eq:triangle-series}
 x_{\rm tri}(t)=\frac{8A}{\pi^2}
 \sum_{m=0}^{\infty}\frac{(-1)^m}{(2m+1)^2}
 \sin[(2m+1)\theta].
\end{equation}
\begin{equation}\label{eq:triangle-frequency}
 X_{\rm obs}^{\rm tri}(\omega)=\frac{4A}{i\pi^2}
 \sum_{m=0}^{\infty}\frac{(-1)^m}{(2m+1)^2}
 \bigl[W(\omega-(2m+1)\omega_0)
       -W(\omega+(2m+1)\omega_0)\bigr].
\end{equation}
The nonzero amplitudes are $A_n=8A/(\pi^2n^2)$ for odd $n$. Square and triangular waves therefore have identical parity selection but different decay: continuity removes the $n^{-1}$ contribution, leaving the $n^{-2}$ tail associated with slope discontinuities.

\subsubsection{Full-wave rectified sine}\label{subsubsec:rectified}

For $x_{\rm r}(t)=A|\sin\theta|$, the fundamental output period is $T/2$. Direct integration gives
\begin{align}
 c_0^{\rm r}
 &=\frac{A}{\pi}\int_0^\pi\sin\theta\,d\theta
 =\frac{2A}{\pi},\notag\\
 c_{2m}^{\rm r}
 &=\frac{A}{\pi}\int_0^\pi
 \sin\theta\cos(2m\theta)\,d\theta
 =-\frac{2A}{\pi(4m^2-1)},\quad m\ge1,
 \label{eq:rectified}\\
 x_{\rm r}(t)
 &=\frac{2A}{\pi}-\frac{4A}{\pi}
 \sum_{m=1}^{\infty}\frac{\cos(2m\theta)}{4m^2-1}.
 \label{eq:rectified-series}
\end{align}
\begin{align}\label{eq:rectified-frequency}
 X_{\rm obs}^{\rm r}(\omega)
 ={}&\frac{2A}{\pi}W(\omega)
 -\frac{2A}{\pi}\sum_{m=1}^{\infty}
 \frac{W(\omega-2m\omega_0)}{4m^2-1}\notag\\
 &-\frac{2A}{\pi}\sum_{m=1}^{\infty}
 \frac{W(\omega+2m\omega_0)}{4m^2-1}.
\end{align}
Relative to the input frequency $\omega_0$, only even indices occur, with peak amplitudes $4A/[\pi(4m^2-1)]\sim A/(\pi m^2)$. The $m^{-2}$ tail follows from cusps at the zeros of the sinusoid. Relative to the output fundamental $2\omega_0$, these are consecutive harmonics.

\subsubsection{Exponential nonlinearity}\label{subsubsec:exponential}

For an exponential response to $q(t)=B\cos\theta$, write
\begin{equation}\label{eq:exp-definition}
 x_{\rm e}(t)=Ae^{\alpha\cos\theta},\qquad
 \alpha=\kappa B,
\end{equation}
where $\kappa$ is an inverse-displacement scale. The modified Bessel functions satisfy
\begin{equation}\label{eq:bessel-I}
 I_n(\alpha)=\frac{1}{\pi}\int_0^\pi
 e^{\alpha\cos\theta}\cos(n\theta)\,d\theta,
\end{equation}
so that
\begin{equation}\label{eq:exp-series}
 x_{\rm e}(t)=AI_0(\alpha)+
 2A\sum_{n=1}^{\infty}I_n(\alpha)\cos(n\theta),
 \qquad c_n^{\rm e}=AI_{|n|}(\alpha).
\end{equation}
\begin{equation}\label{eq:exp-frequency}
 X_{\rm obs}^{\rm e}(\omega)=A\sum_{n\in\mathbb Z}
 I_{|n|}(\alpha)\,W(\omega-n\omega_0).
\end{equation}
For fixed $\alpha\ne0$ and $n\to\infty$,
\begin{equation}\label{eq:exp-asymptotic}
 A_n=2A|I_n(\alpha)|\sim
 \frac{2A}{n!}\left(\frac{|\alpha|}{2}\right)^n.
\end{equation}
The analytic map generates infinitely many nonzero harmonics with factorial large-order decay. A broad spectrum therefore need not arise from waveform discontinuities. For $\alpha<0$, odd-order coefficients change sign without changing their magnitudes.

\begin{figure}
    \centering
    \includegraphics[width=\linewidth]{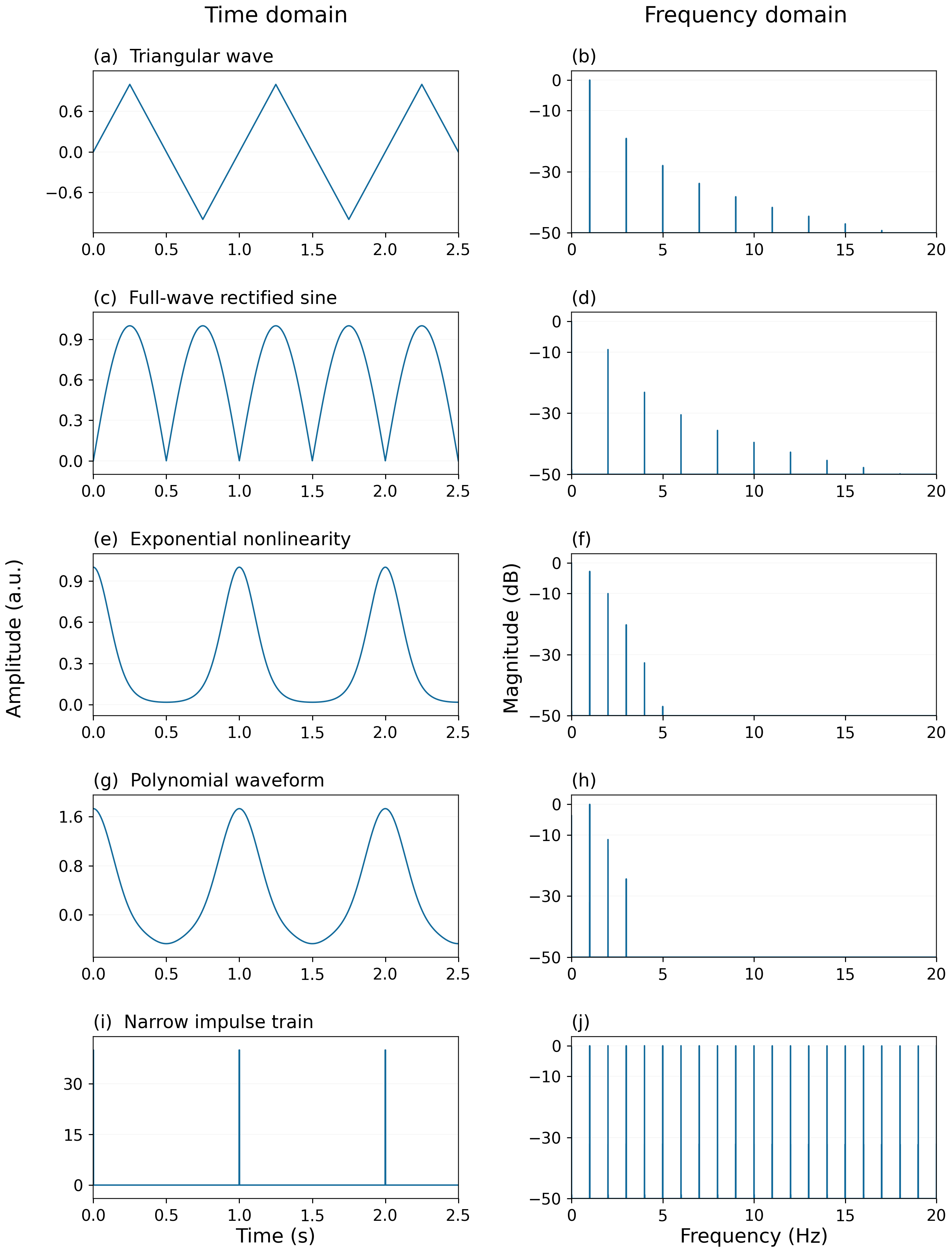}
    \caption{Time-domain waveforms (left column) and their corresponding Fourier magnitude spectra (right column) for five additional mathematical constructions of overtone frequency combs. Panels~(a,b): triangular wave; panels~(c,d): full-wave rectified sine; panels~(e,f): exponential nonlinearity; panels~(g,h): polynomial waveform; and panels~(i,j): narrow impulse train.}
    \label{fig2}
\end{figure}

\subsubsection{Polynomial}\label{subsubsec:polynomial}

Let $F(q)=\sum_{p=0}^P a_pq^p$. Expanding $q(t)=B\cos\theta$ by
\begin{equation}
\label{eq:cos-power}
 \cos^p\theta=2^{-p}\sum_{k=0}^p
 \binom{p}{k}e^{i(p-2k)\theta}
\end{equation}
yields
\begin{equation}\label{eq:poly-coeff}
 c_n^{\rm poly}=\!
 \sum_{\substack{0\le p\le P,\ |n|\le p\\p-n\ \mathrm{even}}}
 \frac{a_pB^p}{2^p}{p\choose (p-n)/2}.
\end{equation}
\begin{align}\label{eq:poly-frequency}
 X_{\rm obs}^{\rm poly}(\omega)
 =\sum_{n=-P}^{P}\!W(\omega-n\omega_0)
 \sum_{\substack{|n|\le p\le P\\p-n\;\mathrm{even}}}
 \frac{a_pB^p}{2^p}\binom{p}{(p-n)/2}.
\end{align}
In particular, a cubic map gives
\begin{align}\label{eq:cubic-output}
 x(t)={}&a_0+\frac{a_2B^2}{2}
 +\left(a_1B+\frac{3a_3B^3}{4}\right)\cos\theta\notag\\
 &+\frac{a_2B^2}{2}\cos2\theta
 +\frac{a_3B^3}{4}\cos3\theta.
\end{align}
Only indices $|n|\le P$ can be generated by a degree-$P$ polynomial applied to a single sinusoid. Even (odd) polynomial terms generate even (odd) indices, with possible cancellation between contributions. This finite-order bound concerns the specified transfer function; it does not bound the harmonic content of a periodic solution to a nonlinear differential equation containing polynomial terms.

\subsubsection{Impulse train}\label{subsubsec:impulse}

The distributional limit of a narrow pulse train is obtained at fixed pulse area $Q=A\tau$:
\begin{equation}\label{eq:dirac-comb}
 x_\delta(t)=Q\sum_{k\in\mathbb Z}\delta(t-kT),
 \qquad c_n^\delta=\frac{Q}{T}\quad(n\in\mathbb Z).
\end{equation}
For rectangular pulses with height $Q/\tau$, Eq.~\eqref{eq:pulse-coeff} becomes
\begin{equation}\label{eq:regularized-comb}
 c_n^{(\tau)}=\frac{Q}{T}
 \operatorname{sinc}\!\left(\frac{\pi n\tau}{T}\right)
 \longrightarrow\frac{Q}{T}\quad(\tau\to0)
\end{equation}
\begin{equation}\label{eq:impulse-frequency}
 X_{\delta,\rm obs}(\omega)=\frac{Q}{T}
 \sum_{n\in\mathbb Z}W(\omega-n\omega_0).
\end{equation}
for each fixed $n$. The ideal coefficients have no decaying envelope, although the limit is not uniform in harmonic order. Finite-width pulses recover a bandwidth set by their width, with the high-order decay determined by edge regularity. The impulse comb is a distribution, not a finite-amplitude mechanical trajectory.

Figures 1 and 2 showcase the time-domain waveforms and the corresponding frequency spectra of overtone frequency combs.

\subsection{Sideband Frequency Combs}\label{sec:sidebands}\label{subsec:full-sidebands}

An overtone family is referenced to a fundamental oscillation and its harmonics, whereas sidebands are naturally referenced to a carrier $\omega_c$. A periodically modulated carrier can generate components at $\omega_c+n\Omega$, where $\Omega$ is the envelope repetition frequency. For an amplitude-modulated signal, write
\begin{equation}\label{eq:am-common}
 x_{\rm AM}(t)=A[1+\mu m(\Omega t)]\cos(\omega_ct),
\end{equation}
where $m(\theta)$ is a real, zero-mean, $2\pi$-periodic waveform and $\mu$ is its modulation depth. With the Fourier expansion
\begin{equation}\label{eq:am-envelope-series}
 m(\theta)=\sum_{n\in\mathbb Z}m_ne^{in\theta},\qquad
 m_n=\frac{1}{2\pi}\int_{-\pi}^{\pi}m(\theta)e^{-in\theta}\,d\theta,
\end{equation}
multiplication by the carrier translates the Fourier lines of $m$ into upper and lower sidebands. In the simulations the carrier frequency is $\omega_c=r\Omega$ with $r=10$. The complete Fourier coefficient of the real signal at index $k$ is then
\begin{align}\label{eq:am-shift}
 c_k^{\rm AM}={}&\frac{A}{2}(\delta_{k,r}+\delta_{k,-r})\notag\\
 &+\frac{A\mu}{2}\left(m_{k-r}+m_{k+r}\right).
\end{align}
The second envelope image cannot always be omitted: when the lower sidebands reach zero frequency or overlap the negative-frequency carrier image, the complex contributions have to be summed. A periodically repeated modulation has frequencies on an integer grid, but symmetry can remove particular orders. To distinguish this selection effect from genuinely uneven observed spacings, we examine consecutive sideband families before turning to nonuniform patterns.

\subsubsection{Sinusoidal frequency or phase modulation}\label{subsubsec:pm}
For a sinusoidal phase excursion $\beta\sin\Omega t$, the signal is
\begin{equation}\label{eq:pm-new}
 x_{\rm PM}(t)=A\cos[\omega_ct+\beta\sin(\Omega t)].
\end{equation}
Using the Jacobi--Anger expansion gives
\begin{equation}\label{eq:pm-series-new}
 x_{\rm PM}(t)=A\sum_{n\in\mathbb Z}J_n(\beta)
 \cos[(\omega_c+n\Omega)t].
\end{equation}
For the real signal at the integer carrier ratio $r$, the complex coefficient is
\begin{equation}\label{eq:pm-coeff-new}
 c_k^{\rm PM}=\frac{A}{2}\bigl[J_{k-r}(\beta)+J_{-k-r}(\beta)\bigr].
\end{equation}
If $\omega_{\rm inst}=\omega_c+\Delta\omega\cos(\Omega t)$, frequency modulation produces the same waveform with $\beta=\Delta\omega/\Omega$. Its sideband weights are governed by ordinary Bessel functions $J_n$. Isolated Bessel zeros can remove individual lines; the equally spaced frequency grid is a structural property, not a claim that every order always has a measurable amplitude \cite{czaplewski2018bifurcation,ochs2022single}.

\begin{figure}
    \centering
    \includegraphics[width=\linewidth]{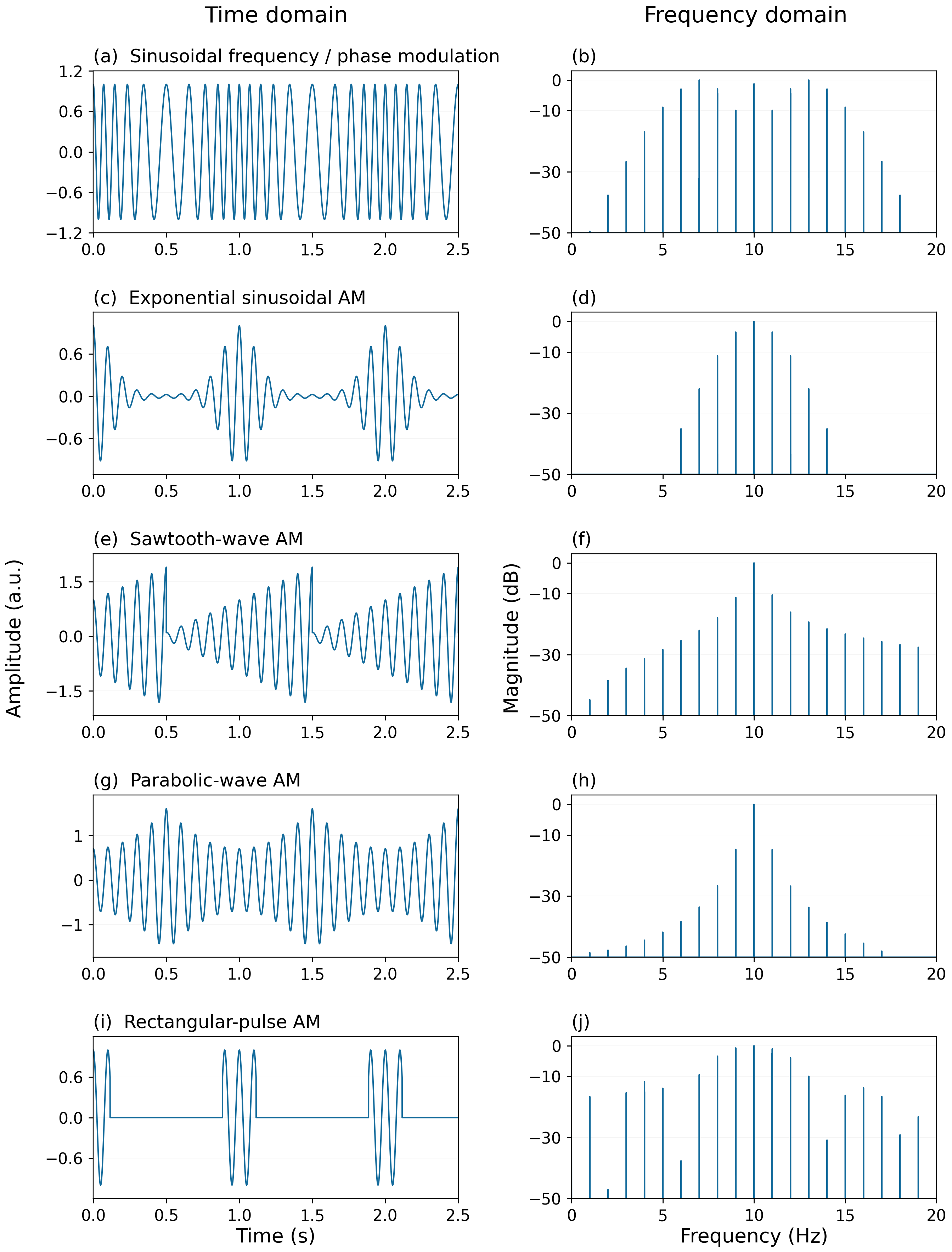}
    \caption{Time-domain signals (left column) and their corresponding Fourier magnitude spectra (right column) for five mathematical constructions of equally spaced sideband frequency combs. Panels~(a,b): sinusoidal frequency or phase modulation; panels~(c,d): exponential sinusoidal amplitude modulation; panels~(e,f): sawtooth-wave amplitude modulation; panels~(g,h): parabolic-wave amplitude modulation; and panels~(i,j): rectangular-pulse amplitude modulation.}
    \label{fig3}
\end{figure}

\subsubsection{Exponential sinusoidal amplitude modulation}\label{subsubsec:exp-am}
A positive, smooth envelope can be formed from a sinusoid:
\begin{equation}\label{eq:exp-am}
 x_{\rm eAM}(t)=Ae^{\eta[\cos(\Omega t)-1]}\cos(\omega_ct),
\end{equation}
where $\eta>0$ controls the envelope concentration. The modified-Bessel identity
\begin{equation}\label{eq:exp-am-bessel}
 e^{\eta\cos\theta}=\sum_{n\in\mathbb Z}I_{|n|}(\eta)e^{in\theta}
\end{equation}
gives the exact coefficients
\begin{equation}\label{eq:exp-am-coeff}
 c_k^{\rm eAM}=\frac{Ae^{-\eta}}{2}
 \bigl[I_{|k-r|}(\eta)+I_{|k+r|}(\eta)\bigr].
\end{equation}
Every envelope order is allowed for $\eta>0$, while the large-order amplitudes decrease factorially. This contrasts with envelopes possessing sharp corners or discontinuities.

\subsubsection{Sawtooth-wave amplitude modulation}\label{subsubsec:saw-am-new}
Let $u(\theta)$ denote the representative of $\theta$ in $[-\pi,\pi)$. The normalized ascending sawtooth is $m_{\rm saw}(\theta)=u(\theta)/\pi$, so
\begin{equation}\label{eq:saw-am-new}
 x_{\rm saw,AM}(t)=A\left[1+\mu_{\rm saw}\frac{u(\Omega t)}{\pi}\right]
 \cos(\omega_ct).
\end{equation}
Its complex envelope coefficients are
\begin{equation}\label{eq:saw-am-coeff-new}
 m_n^{\rm saw}=\frac{(-1)^{|n|+1}}{i\pi n}\quad(n\ne0),
 \qquad m_0^{\rm saw}=0.
\end{equation}
All nonzero positive orders appear, with a $1/n$ envelope due to the discontinuous periodic reset. Their exact carrier-centred amplitudes follow from Eq.~\eqref{eq:am-shift}.

\subsubsection{Parabolic-wave amplitude modulation}\label{subsubsec:parabolic-am}
The periodic parabola introduced in Sec.~\ref{subsec:periodic-parabolic} gives another elementary envelope:
\begin{equation}\label{eq:par-am}
 x_{\rm B,AM}(t)=A\left[1+\mu_{\rm B}
 \left(\frac{u(\Omega t)^2}{\pi^2}-\frac13\right)\right]
 \cos(\omega_ct).
\end{equation}
Here
\begin{equation}\label{eq:par-am-coeff}
 m_0^{\rm B}=0,\qquad
 m_{\pm n}^{\rm B}=\frac{2(-1)^n}{\pi^2 n^2}\quad(n\ge1).
\end{equation}
No harmonic order is forbidden, but the envelope is smoother than the sawtooth and its sidebands decay as $n^{-2}$ away from overlapping carrier images.

\subsubsection{Rectangular-pulse amplitude modulation}\label{subsubsec:pulse-am-new}
For a centred periodic gate of duty cycle $d$, let
\begin{equation}\label{eq:gate-def-new}
 g_d(\theta)=\begin{cases}1,&|u(\theta)|<\pi d,\\0,&\text{otherwise},\end{cases}
 \qquad 0<d<1.
\end{equation}
The gated carrier is $x_{\rm g,AM}(t)=Ag_d(\Omega t)\cos\omega_ct$. Direct integration yields
\begin{equation}\label{eq:gate-envelope}
 (g_d)_n=d\,\operatorname{sinc}(\pi nd),
\end{equation}
with $\operatorname{sinc}z=\sin z/z$. Consequently,
\begin{align}\label{eq:gate-coeff-new}
 c_k^{\rm g,AM}=\frac{Ad}{2}\bigl[&\operatorname{sinc}(\pi(k-r)d)\notag\\
 &+\operatorname{sinc}(\pi(k+r)d)\bigr].
\end{align}
The spacing is $\Omega$, although exact sinc nulls may eliminate orders for special duty cycles. The value $d=0.23$ used in the plotted example avoids these low-order envelope nulls in the displayed frequency interval.

Figure 3 showcases the time-domain waveforms and the corresponding frequency spectra of sideband frequency combs.

\subsection{Nonuniformly populated sideband spectra}\label{subsec:sparse-sidebands}

We now consider waveforms that do not produce equidistant frequency comb spectra. Square and triangular envelopes permit only odd envelope harmonics, which are \emph{missing orders on an equally spaced grid}, not intrinsically irregular frequency positions. A two-tone envelope with incommensurate frequencies does not possess single modulation-frequency grid. A finite superposition with quadratic harmonic indices remains periodic, but its populated lines have unequal successive gaps. The separate constructions below make these distinctions explicit.

\subsubsection{Square-wave amplitude modulation}\label{subsubsec:square-am-new}
The square-wave envelope is
\begin{equation}\label{eq:square-am-new}
 m_{\rm sq}(\theta)=\operatorname{sgn}(\sin\theta),\qquad
 x_{\rm sq,AM}(t)=A[1+\mu m_{\rm sq}(\Omega t)]
 \cos(\omega_ct).
\end{equation}
Its coefficients are
\begin{equation}\label{eq:square-env-new}
 m_n^{\rm sq}=\begin{cases}
 2/(i\pi n),&n\ \mathrm{odd},\\
 0,&n\ \mathrm{even}.
 \end{cases}
\end{equation}
Thus sidebands associated with even envelope orders are absent. The allowed offsets are odd multiples of $\Omega$, spaced by $2\Omega$ within each sideband wing, with amplitudes decreasing as $1/|n|$.

\subsubsection{Triangular-wave amplitude modulation}\label{subsubsec:tri-am-new}
For a continuous piecewise-linear envelope,
\begin{equation}\label{eq:tri-am-new}
 m_{\rm tri}(\theta)=\frac{2}{\pi}\arcsin(\sin\theta),\quad
 x_{\rm tri,AM}=A[1+\mu m_{\rm tri}(\Omega t)]\cos\omega_ct.
\end{equation}
The coefficients are
\begin{equation}\label{eq:tri-env-new}
 m_n^{\rm tri}=\begin{cases}
 \dfrac{4(-1)^{(|n|-1)/2}}{i\pi^2 n|n|},&n\ \mathrm{odd},\\[2pt]
 0,&n\ \mathrm{even}.
 \end{cases}
\end{equation}
The same odd-only selection rule holds, but the continuous waveform has a $1/n^2$ rather than a $1/|n|$ spectral envelope.

\begin{figure}
    \centering
    \includegraphics[width=\linewidth]{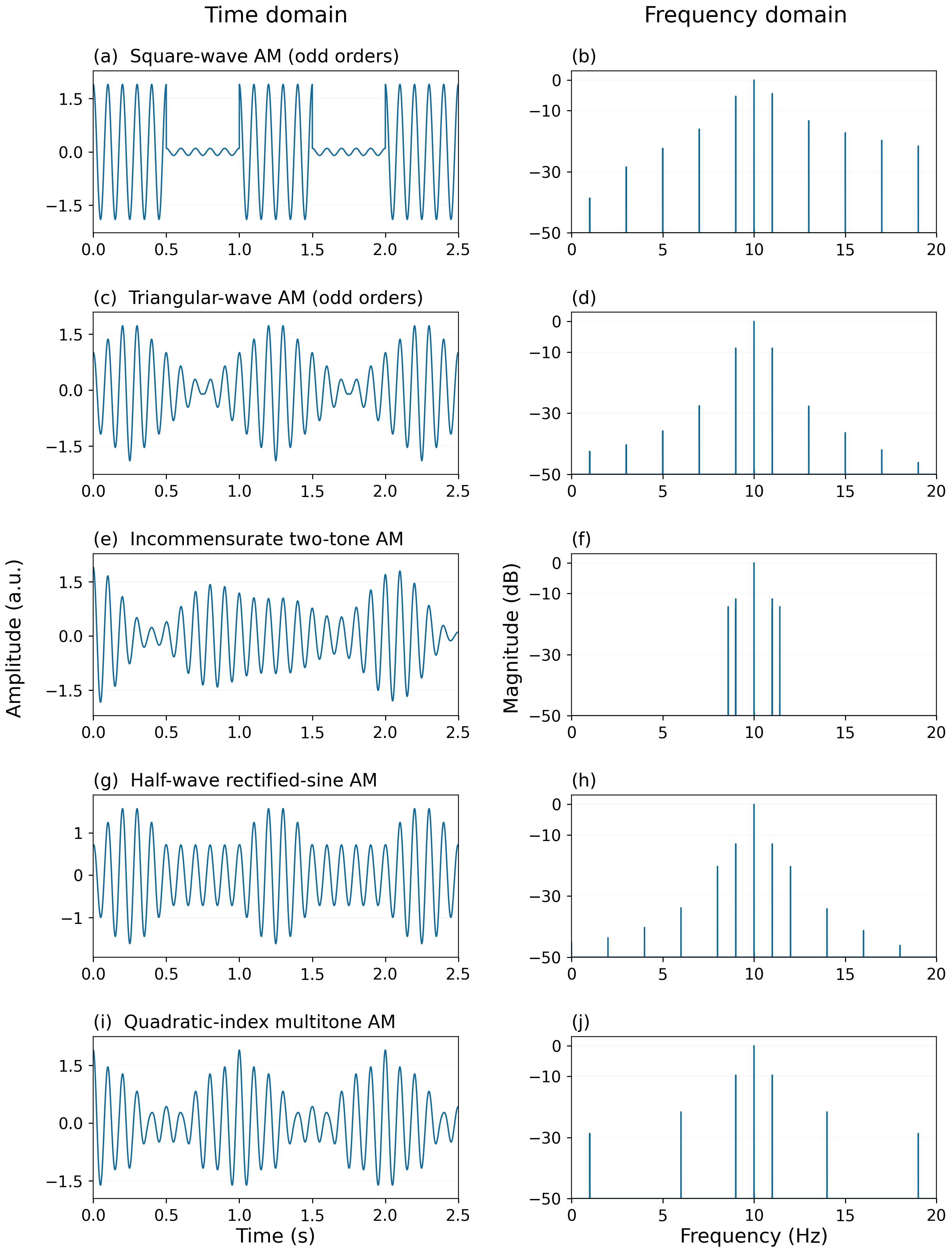}
    \caption{Time-domain signals (left column) and their corresponding Fourier magnitude spectra (right column) for five mathematical constructions producing nonuniform sideband distributions. Panels~(a,b): square-wave amplitude modulation; panels~(c,d): triangular-wave amplitude modulation; panels~(e,f): incommensurate two-tone amplitude modulation; panels~(g,h): half-wave rectified-sine amplitude modulation; and panels~(i,j): quadratic-index multitone amplitude modulation.}
    \label{fig4}
\end{figure}

\subsubsection{Incommensurate two-tone amplitude modulation}\label{subsubsec:two-tone-am}
Two simple sinusoidal modulations with an irrational frequency ratio provide a genuine example without a common equidistant frequency lattice. Set $\rho=\sqrt{2}$ and $b>0$, and write
\begin{equation}\label{eq:two-tone-am}
 x_{\rm 2t}(t)=A\left[1+\frac{\mu}{1+b}
 \bigl(\cos\Omega t+b\cos\rho\Omega t\bigr)\right]
 \cos\omega_ct.
\end{equation}
The product-to-sum identity yields
\begin{align}\label{eq:two-tone-series}
 x_{\rm 2t}(t)={}&A\cos\omega_ct\notag\\
 &+\frac{A\mu}{2(1+b)}
 \sum_{\sigma=\pm1}\cos[(\omega_c+\sigma\Omega)t]\notag\\
 &+\frac{A\mu b}{2(1+b)}
 \sum_{\sigma=\pm1}\cos[(\omega_c+\sigma\rho\Omega)t].
\end{align}
The four sidebands are at $\omega_c\pm\Omega$ and $\omega_c\pm\sqrt2\Omega$. Their successive positive-frequency gaps are unequal, and the complete waveform has no common period. This example produces a few \emph{nonuniformly spaced sidebands}, rather than a frequency comb.

\subsubsection{Half-wave rectified-sine amplitude modulation}\label{subsubsec:halfrect-am}
Let $h(\theta)=\max(0,\sin\theta)$ and subtract its mean to obtain
\begin{equation}\label{eq:halfrect-def}
 m_{\rm h}(\theta)=h(\theta)-\frac1\pi,\qquad
 x_{\rm h,AM}(t)=A[1+\mu m_{\rm h}(\Omega t)]
 \cos\omega_ct.
\end{equation}
Its Fourier expansion contains a first sine harmonic and even cosine harmonics:
\begin{equation}\label{eq:halfrect-series}
 m_{\rm h}(\theta)=\frac12\sin\theta
 -\frac{2}{\pi}\sum_{j=1}^{\infty}
 \frac{\cos(2j\theta)}{4j^2-1}.
\end{equation}
Equivalently, $m_{1}^{\rm h}=1/(4i)$, $m_{-1}^{\rm h}=-1/(4i)$, $m_{\pm 2j}^{\rm h}=-1/[\pi(4j^2-1)]$, and all other coefficients vanish. The allowed carrier offsets contain $\pm\Omega$ and $\pm2j\Omega$, but not higher odd indices. Its observed orders therefore have gaps which alternate as one proceeds away from the carrier.

\subsubsection{Quadratic-index multitone amplitude modulation}\label{subsubsec:quadratic-am}
A second explicit example of nonuniform gaps follows from a finite sum of familiar cosine functions. For $p>0$, define
\begin{equation}\label{eq:quadratic-envelope}
 m_{\rm q}(\theta)=\frac{1}{S_p}
 \sum_{j=1}^{3}\frac{\cos(j^2\theta)}{j^p},\qquad
 S_p=\sum_{j=1}^{3}j^{-p},
\end{equation}
and take
\begin{equation}\label{eq:quadratic-am}
 x_{\rm q,AM}(t)=A[1+\mu m_{\rm q}(\Omega t)]
 \cos\omega_ct.
\end{equation}
Applying the product-to-sum identity gives the finite sideband spectrum
\begin{align}\label{eq:quadratic-series}
 x_{\rm q,AM}(t)={}&A\cos\omega_ct
 +\frac{A\mu}{2S_p}\sum_{j=1}^{3}j^{-p}\notag\\
 &\times\left\{\cos[(\omega_c+j^2\Omega)t]
 +\cos[(\omega_c-j^2\Omega)t]\right\}.
\end{align}
The carrier-relative offsets are $\Omega$, $4\Omega$, and $9\Omega$. Their successive gaps are $3\Omega$ and $5\Omega$, so the \emph{occupied} sideband frequencies are not consecutive or uniformly separated. Nevertheless, unlike the incommensurate two-tone case, these frequencies remain subsets of the ordinary integer $\Omega$ grid. The simulation uses $p=2$.

Figure 4 showcases the time-domain waveforms and their corresponding nonuniform frequency spectra.

\section{Conclusion}

We have established a Fourier-based framework connecting time-domain waveform properties to the spectral structure of overtone and sideband frequency combs. For periodic waveforms, symmetry determines the allowed harmonic orders, whereas regularity governs their spectral decay. Discontinuities produce $n^{-1}$ amplitude scaling, derivative discontinuities yield $n^{-2}$ scaling, and smooth nonlinear mappings can generate more rapidly decaying spectra. These analytical results demonstrate how distinct waveforms can produce equally spaced harmonics with markedly different amplitude distributions. For modulated carriers, the analysis distinguishes consecutive sidebands, symmetry-induced missing orders, and nonuniformly spaced frequency components. In particular, spectral sparsity does not necessarily imply a loss of periodicity or the absence of an underlying uniform frequency lattice. These findings provide mathematical criteria for interpreting frequency-comb spectra without attributing them to a particular mechanism.

\bibliographystyle{unsrt}  
\bibliography{references}  

@article{udem2002optical,
  author={Udem, Theodor and Holzwarth, Ronald and H{\"a}nsch, Theodor W.},
  title={Optical frequency metrology}, journal={Nature},
  volume={416}, pages={233--237}, number={6877}, year={2002}, doi={10.1038/416233a}
}

@article{cundiff2003colloquium,
  author={Cundiff, Steven T. and Ye, Jun},
  title={Colloquium: Femtosecond optical frequency combs},
  journal={Reviews of Modern Physics}, volume={75}, pages={325--342}, number={1}, year={2003},
  doi={10.1103/RevModPhys.75.325}
}

@article{fortier2019twenty,
  author={Fortier, Tara and Baumann, Esther},
  title={20 years of developments in optical frequency comb technology and applications},
  journal={Communications Physics}, volume={2}, pages={153}, year={2019},
  doi={10.1038/s42005-019-0249-y}
}

@article{mahboob2012tuneable,
  author={Mahboob, I. and Wilmart, Q. and Nishiguchi, K. and Fujiwara, A. and Yamaguchi, H.},
  title={Tuneable electromechanical comb generation},
  journal={Applied Physics Letters}, volume={100}, pages={113109}, number={11}, year={2012},
  doi={10.1063/1.3694041}
}

@article{cao2014phononic,
  author={Cao, L. S. and Qi, D. X. and Peng, R. W. and Wang, Mu and Schmelcher, P.},
  title={Phononic frequency combs through nonlinear resonances},
  journal={Physical Review Letters}, volume={112}, pages={075505}, number={7}, year={2014},
  doi={10.1103/PhysRevLett.112.075505}
}

@article{ganesan2017phononic,
  author={Ganesan, Adarsh and Do, Cuong and Seshia, Ashwin},
  title={Phononic frequency comb via intrinsic three-wave mixing},
  journal={Physical Review Letters}, volume={118}, pages={033903}, number={3}, year={2017},
  doi={10.1103/PhysRevLett.118.033903}
}

@article{ganesan2017fourwave,
  author={Ganesan, Adarsh and Do, Cuong and Seshia, Ashwin},
  title={Frequency transitions in phononic four-wave mixing},
  journal={Applied Physics Letters}, volume={111}, number={6}, pages={064101}, year={2017},
  doi={10.1063/1.4985266}
}

@article{ganesan2018threemode,
  author={Ganesan, Adarsh and Do, Cuong and Seshia, Ashwin},
  title={Phononic frequency comb via three-mode parametric resonance},
  journal={Applied Physics Letters}, number={2}, pages={021906}, year={2018},
  doi={10.1063/1.5003133}
}

@article{czaplewski2018bifurcation,
  author={Czaplewski, David A. and Chen, Changyao and Lopez, Daniel and Shoshani, Oriel and Eriksson, Axel M. and Strachan, Scott and Shaw, Steven W.},
  title={Bifurcation generated mechanical frequency comb},
  journal={Physical Review Letters}, volume={121}, pages={244302}, number={24}, year={2018},
  doi={10.1103/PhysRevLett.121.244302}
}

@article{seitner2017parametric,
  author={Seitner, Maximilian J. and Abdi, Mehdi and Ridolfo, Alessandro and Hartmann, Michael J. and Weig, Eva M.},
  title={Parametric oscillation, frequency mixing, and injection locking of strongly coupled nanomechanical resonator modes},
  journal={Physical Review Letters}, volume={118}, pages={254301}, number={25}, year={2017},
  doi={10.1103/PhysRevLett.118.254301}
}

@article{goryachev2020generation,
  author={Goryachev, Maxim and Galliou, Serge and Tobar, Michael E.},
  title={Generation of ultralow power phononic combs},
  journal={Physical Review Research}, volume={2}, pages={023035}, number={2}, year={2020},
  doi={10.1103/PhysRevResearch.2.023035}
}

@article{ochs2022single,
  author={Ochs, J. S. and Bone{\ss}, D. K. J. and Rastelli, G. and Seitner, M. and Belzig, W. and Dykman, M. I. and Weig, E. M.},
  title={Frequency comb from a single driven nonlinear nanomechanical mode},
  journal={Physical Review X}, volume={12}, pages={041019}, number={4}, year={2022},
  doi={10.1103/PhysRevX.12.041019}
}

@article{keskekler2022symmetry,
  author={Ke{\c{s}}kekler, Ata and Arjmandi-Tash, Hadi and Steeneken, Peter G. and Alijani, Farbod},
  title={Symmetry-breaking-induced frequency combs in graphene resonators},
  journal={Nano Letters}, volume={22}, number={15}, pages={6048--6054}, year={2022},
  doi={10.1021/acs.nanolett.2c00360}
}

@article{batista2020frequency,
  author={Batista, Adriano A. and {Lisboa de Souza}, A. A.},
  title={Frequency-comb response of a parametrically driven {Duffing} oscillator to a small added ac excitation},
  journal={Journal of Applied Physics}, volume={128}, pages={244901}, number={24}, year={2020},
  doi={10.1063/5.0029104}
}

@article{dejong2023mechanical,
  author={de Jong, Matthijs H. J. and Ganesan, Adarsh and Cupertino, Andrea and Gr{\"o}blacher, Simon and Norte, Richard A.},
  title={Mechanical overtone frequency combs},
  journal={Nature Communications}, volume={14}, pages={1458}, year={2023},
  doi={10.1038/s41467-023-36953-8}
}

@article{bhosale2024multiharmonic,
  author={Bhosale, Kalyani S. and Li, Sheng-Shian},
  title={Multi-harmonic phononic frequency comb generation in capacitive {CMOS-MEMS} resonators},
  journal={Applied Physics Letters}, volume={124}, pages={163505}, number={16}, year={2024},
  doi={10.1063/5.0197773}
}

@article{miri2018optomechanical,
  author={Miri, Mohammad-Ali and D'Aguanno, Giuseppe and Al{\`u}, Andrea},
  title={Optomechanical frequency combs},
  journal={New Journal of Physics}, volume={20}, pages={043013}, number={4}, year={2018},
  doi={10.1088/1367-2630/aab5c6}
}

@article{wu2022hybridized,
  author={Wu, Sishi and Liu, Yulong and Liu, Qichun and Wang, Shuai-Peng and Chen, Zhen and Li, Tiefu},
  title={Hybridized frequency combs in multimode cavity electromechanical system},
  journal={Physical Review Letters}, volume={128}, pages={153901}, number={15}, year={2022},
  doi={10.1103/PhysRevLett.128.153901}
}

@article{xiao2026ultrabroadband,
  author={Xiao, Guangzong and Feng, Zijian and Kuang, Tengfang and Huang, Ran and He, Yutong and Chen, Xinlin and Zuo, Yunlan and Han, Xiang and Xiong, Wei and Tan, Zhongqi and Ganesan, Adarsh and Nori, Franco and Qiu, Cheng-Wei and Zhang, Xiaobao and Luo, Hui and Jing, Hui},
  title={Ultrabroadband phonon laser frequency comb},
  journal={Advanced Photonics}, volume={8}, number={2}, pages={026004}, year={2026},
  doi={10.1117/1.AP.8.2.026004}
}

@misc{NISTDLMF,
  author={{NIST Digital Library of Mathematical Functions}},
  title={{DLMF} Chapter 22: Jacobian Elliptic Functions, Section 22.11: Fourier and Hyperbolic Series},
  howpublished={National Institute of Standards and Technology},
  note={Accessed 2026-10-27}, url={https://dlmf.nist.gov/22.11}
}
\end{document}